\documentclass[prc]{revtex4}

\usepackage{graphicx,amsmath}

\begin{document}
\title{Removing Distortions from Charge Balance Functions}

\author{Scott Pratt}
\email{pratt@nscl.msu.edu}
\affiliation{Department of Physics and National Superconducting Cyclotron
Laboratory\\
Michigan State University, East Lansing Michigan, 48824}

\author{Sen Cheng}
\altaffiliation[current address: ]
{Sloan-Swartz Center, University of California at San Francisco\\
513 Parnassus Ave., Box 0444\\
San Francisco, CA 94143-0444, USA}
\affiliation{Department of Physics and Astronomy and National Superconducting 
Cyclotron Laboratory\\
Michigan State University, East Lansing Michigan, 48824}

\date{\today}

\begin{abstract}
\bigskip
Charge balance functions provide insight into critical issues concerning
hadronization and transport in heavy-ion collisions by statistically isolating
charge/anti-charge pairs which are correlated by charge conservation. However,
distortions from residual interactions and unbalanced charges cloud the
observable. Within the context of simple models, the significance of these
effects is studied by constructing balance functions in both relative rapidity
and invariant relative momentum. Methods are presented for eliminating or
accounting for these distortions.

\end{abstract}

\maketitle

\section{Introduction}
\label{sec:intro}
Charge balance functions were suggested as a means for addressing fundamental
questions concerning hadronization in relativistic heavy ion collisions
\cite{bassdanpratt}. The most pressing issue concerns whether hadronization is
delayed in such reactions beyond the characteristic time scale of 1 fm/c, i.e.,
is a new phase of matter created?  A delayed hadronization of a gluon-rich
medium would mean that many charge-anticharge pairs would be created late in
the reaction and then be more tightly correlated to one another in momentum
space. Charge balance functions are designed to identify such charge/anticharge
pairs on a statistical basis. Unfortunately, the ability to identify balancing
partners is compromised by two effects. First, surplus charge, originating from
the non-zero baryon number and charge of the colliding nuclei, pollutes the
balance function. Secondly, interactions of a balancing pair with the other
charges effectively polarize the other particles and distort the shape of the
balance function. In this paper, the magnitude of such distortions are
exhibited within the context of simple blast-wave models, and means for
eliminating or reducing these distortions are presented.

Charge balance functions are based on conditional distributions,
\begin{eqnarray}
\label{eq:balancedef}
B(P_2|P_1)&\equiv&\frac{1}{2}\left\{
\frac{N_{+-}(P_1,P_2)-N_{++}(P_1,P_2)}{N_+(P_1)}
+\frac{N_{-+}(P_1,P_2)-N_{--}(P_1,P_2)}{N_-(P_1)}
\right\}.
\end{eqnarray}
Here, $N_{ab}(P_1,P_2)$ counts the incidences where a particle of charge $a$ is
observed with momentum in a region defined by $P_1$ while a particle of charge
$b$ is observed that satisfies the momentum constraint $P_2$. The second
constraint could be any function of the momenta of the two particles, e.g., the
relative rapidity. Put into words, the balance function measures the chance of
observing an extra particle of opposite charge given the observation of the
first particle. Balance functions are related to charge fluctuations which can
be used to investigate similar issues
\cite{jeonpratt,asakawa,asakawaprl,kochpolonica,kochqm2001,bleicherjeonkoch,
jeonkoch,gavinqm2002,pruneaugavinvoloshin,gavinkapusta}. The advantage of
balance functions is that they represent a more differential measure.

For a neutral system, every charge has an opposite balancing charge and the
balance function would integrate to unity.
\begin{equation}
\sum_{P_2}B(P_2|P_1)=1.
\end{equation}
The normalization is reduced if not all particles carrying the charge are
included, e.g. only $\pi^+$ and $\pi^-$ are evaluated, thus neglecting the
chance that the electric charge is balanced by a kaon or a baryon, or that the
detector has less than a perfect acceptance. If $P_2$ refers to the relative
rapidity, and $P_1$ includes all measured particles, $B(P_2=\Delta Y)$ provides
the probability that a balancing charge was observed with relative rapidity
$\Delta Y$. Since much of the charge observed in a relativistic heavy ion
collision should be produced at hadronization, a delayed hadronization should
result in a tighter space-time correlation between balancing charges. Due to
the large collective flow fields in these reactions, a tighter correlation in
space-time translates into a tighter correlation between the final
momenta. Therefore, a delayed hadronization should be signaled by a narrower
balance function when plotted as a function of relative momentum or relative
rapidity.

One of the most enticing results from RHIC is the preliminary measurement of
balance functions by the STAR collaboration \cite{ray_qm2002star}. In
accordance with expectations for delayed hadronization, the balance functions
appear to narrow with increasing centrality of the collision. However, given
the nascent stage of these observations and of the phenomenology, it should be
emphasized that numerous questions remain concerning the interpretation of such
a measurement. To that end, several issues were pursued in a previous paper,
including the effects of Hanbury-Brown Twiss correlations, detector acceptance
and the relation to charge fluctuations \cite{jeonpratt}.

In the same spirit as that paper, more issues will be addressed in this
study. In the next section, the benefits analyzing balance functions in other
observables, e.g, the invariant relative momentum, will be addressed.  In
addition to allowing one to analyze the contribution from specific resonances,
it will be shown that such observables help clarify other issues such as the
interplay of collective flow and cooling.

Balance function analyses are based on the assumption that all charges have
balancing partners. This is not true in relativistic heavy ion collisions due
to the presence of the initial protons and neutrons which bring about an
imbalance of baryon number, electric charge and isospin. In section
\ref{sec:surplus}, the distorting influence of the surplus positive charge is
investigated and a modified balance function observable is proposed that would
eliminate such effects.

The subsequent section contains a detailed study of the effects of inter-pair
correlations. By extending the model presented in \cite{jeonpratt} to balance
functions in $Q_{\rm inv}$, it appears that the Hanbury-Brown Twiss (HBT)
correlations cause a more noticeable distortion, especially in the most central
collisions. The source of these residual effects is analyzed in detail, and the
degree to which these distortions can be accounted for is discussed.  The final
section presents a summary of what further work must be done in analyzing and
interpreting this class of observables.

\section{Analyzing the Balance Function in $Q_{\rm inv}$}
\label{sec:qinv}

In reference \cite{bassdanpratt} balance functions were evaluated as a function
of relative rapidity. Like two-particle correlation functions, the balance
function is a six-dimensional quantity and new insights can be gained by
performing different cuts or binnings. Specifically, we focus on performing
analyses in terms of the invariant relative momentum, i.e. the relative
momentum as measured by an observer moving with the velocity of the
two-particle center of mass. We find that these variables yield clearer insight
for interpreting the physics of the balancing charges, as well as providing a
better illumination of the distorting effects which are the subject of this
study.

The relative momentum of the two particles is
defined,
\begin{eqnarray}
q_\alpha&\equiv&
(p_{a,\alpha}-p_{b,\alpha})-P_\alpha\frac{P\cdot(p_a-p_b)}{P^2}\\
&=&(p_{a,\alpha}-p_{b,\alpha})-P_\alpha\frac{(m_a^2-m_b^2)}{s}.
\nonumber
\end{eqnarray}
Here, the total momentum of the pair is $P$ and the center-of-mass energy of
the pair is $\sqrt{s}=\sqrt{(p_a+p_b)^2}$. For two particles of the same mass,
the last term can be neglected. The invariant momentum is then
\begin{equation}
Q_{\rm inv}^2=-q^2=-(p_a-p_b)^2+\frac{(m_a^2-m_b^2)^2}{P^2}.
\end{equation}

For pion-correlation studies, it is conventional to define three projections of
the relative momentum, $Q_{\rm long}$, $Q_{\rm out}$ and $Q_{\rm side}$
\cite{prattprd86,bertschoutlongside,csorgopratt}. These components measure the
projections of $q$ along the beam axis, the outwards direction (defined by the
pair's transverse momentum) and the sidewards direction (perpendicular to the
pair's transverse momentum and to the beam axis). Motivated by the semi-boost
invariant nature of the collision geometry at RHIC, $Q_{\rm long}$ is usually
measured in a reference frame moving with the beam velocity of the
pair. Although not typically invoked in correlations studies, one can also
perform a second outwards boost to a frame where the total spatial momentum of
the pair is zero. In this frame the three components, $Q_{\rm out}$, $Q_{\rm
side}$ and $Q_{\rm long}$ sum to $Q_{\rm inv}$.
\begin{equation}
Q_{\rm long}^2+Q_{\rm side}^2+Q_{\rm out}^2=Q_{\rm inv}^2.
\end{equation}
In terms of laboratory momenta $P$ and $q$, these components are:

\begin{eqnarray}
Q_{\rm long}&=&\frac{1}{\sqrt{s+P_t^2}}(P_0q_z-P_zq_0)\\
\nonumber
Q_{\rm side}&=&\frac{(P_xq_y-P_yq_x)}{P_t}\\
\nonumber
 Q_{\rm out}&=&\sqrt{\frac{s}{s+P_t^2}}\frac{(P_xq_x+P_yq_y)}{P_t}.
\end{eqnarray}
Here, $P_t=[(p_{a,x}+p_{b,x})^2+(p_{a,y}+p_{b,y})]^{1/2}$ is the transverse
momentum of the pair. These components differ from the common convention for
HBT in that $Q_{\rm out}$ is defined as the relative momentum in the pair
frame, whereas in HBT the usual convention is to ignore the second boost which
means that the three components do not sum to $Q_{\rm inv}$. In fact, the
factor $\sqrt{(s+P_t^2)/s}$ in the definition of $Q_{\rm out}$ is simply the
Lorentz gamma factor corresponding to the transverse boost to the two-particle
rest frame.

Analyzing balance function in terms of $Q_{\rm inv}$ simplifies interpretation
with thermal models by eliminating the sensitivity to collective flow. Blast
wave models are based on thermal emission from sources which move to account
for the collective flow of the exploding matter. Collective flow affects the
spectra, but leaves the invariant momentum differences unchanged if the two
particles originate from the same space-time point of the blast wave. Hence,
plotting the balance function in invariant momentum variables would minimize
the confusion associated with the collective flow as the width would only
depend on the local thermal properties of the individual sources. If a particle
and its balancing particle were always emitted close to one another in
coordinate space, the width of the balance function would principally be a
function of the breakup temperature with no sensitivity to collective flow,
assuming a uniform detector acceptance.

To illustrate the complications of using rapidity differences rather than
$Q_{\rm inv}$, one may consider a thermal source where the width in $Q_{\rm
long}$ is determined by the temperature. The separation of two tracks in
rapidity is then,
\begin{equation}
\Delta y\sim \frac{Q_{\rm long}}{m_t},
\end{equation}
where $m_t$ is the transverse mass of the particles. Since collective flow
affects the distribution of transverse masses, the balance function widths for
localized thermal sources would depend on the collective flow in the data when
plotted in relative rapidity. Although it is easy to account for collective
flow in a theoretical model, the interpretation of experimental results is
simplified by performing the analysis in $Q_{\rm inv}$.

Furthermore, assuming thermal emission with highly localized charge
conservation, the balance function would be isotropic with respect to the
direction of the relative momentum, e.g., the width in $Q_{\rm side}$ would
equal the width in $Q_{\rm long}$. For early production of charge, one expects
string dynamics or diffusion to lead to an anisotropic balance function as the
balancing charges should separate significantly in coordinate space along the
beam axis due to the extremely large velocity gradient along the beam axis at
early times, $dv_z/dz=1/\tau$. Thus, in addition to the width of the balance
function in $Q_{\rm inv}$, the behavior of the anisotropy as a function of the
collision's centrality provides a crucial test of the mechanism for charge
creation and transport.

\begin{figure}
\centerline{\includegraphics[width=0.7\textwidth]{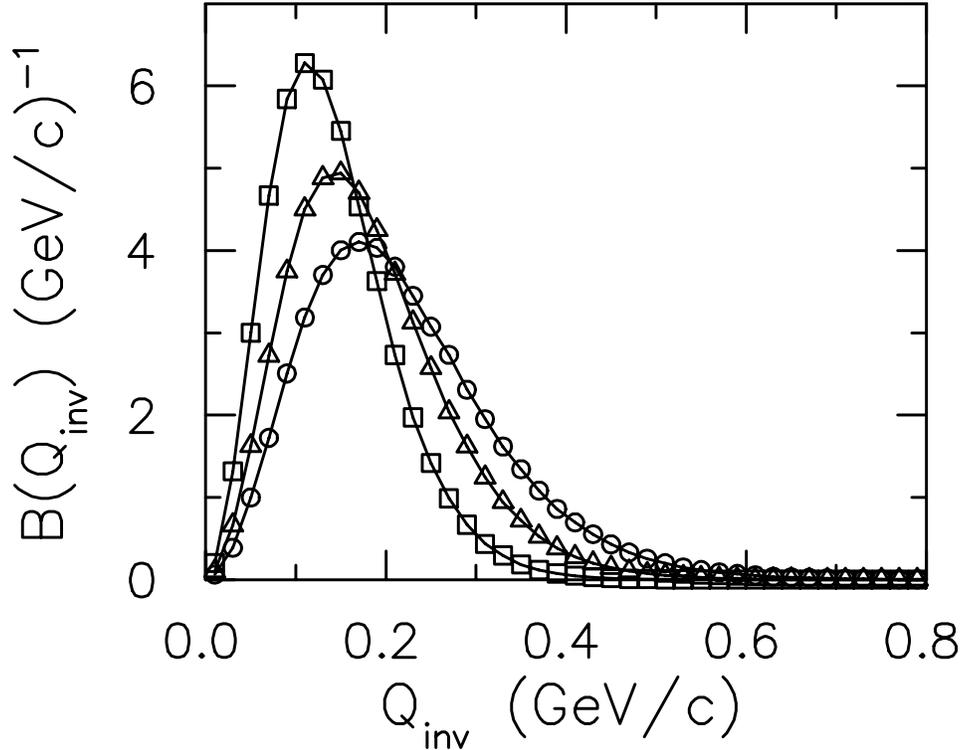}}
\caption{
\label{fig:tempdependence}
Blast-wave predictions of $\pi^+\pi^-$ balance functions are displayed for
three temperatures, assuming the balancing pions are always emitted thermally
from sources with identical source velocities. When plotted in $Q_{\rm inv}$,
the shape depends only on the breakup temperature. Calculations are shown for
$T=90$ (squares), $T=120$ (triangles) and $T=150$ MeV (circles).  }
\end{figure}

To illustrate the sensitivity of a balance function in terms of these analyses,
we consider a simple blast wave model where the collective transverse motion is
assumed to rise linearly with the radius. Of the numerous parameterizations of
the blast wave model, it is assumed that the sources have transverse rapidities
governed by a simple distribution,
\begin{equation}
\frac{dN}{y_t dy_t}=\left\{\begin{array}{cc}
{\rm constant},& y_t<\tanh^{-1}(v_{\rm max})\\
0,&y_t>\tanh^{-1}(v_{\rm max})\\
\end{array}\right.
\end{equation}
Here, $y_t$ is the transverse rapidity, $\tanh(y_t)=v_\perp$. The distribution
of longitudinal rapidities is assumed to be uniform. For our calculation we
assume that these sources emit isotropically in the source frame according to a
temperature $T$. A balancing positive and negative pion are assumed to be
emitted from sources with the same longitudinal and transverse
rapidity. Figure \ref{fig:tempdependence} illustrates the sensitivity to the
temperature by presenting balance functions for three temperatures, 90 MeV, 120
MeV and 150 MeV. The balance function is clearly narrower for lower breakup
temperatures. The results of Fig. \ref{fig:tempdependence} are insensitive to
the choice of $v_{\rm max}$. However, it must be stressed that the sensitivity
would return if the balance function is analyzed in a finite acceptance.

\begin{figure}
\centerline{\includegraphics[width=0.7\textwidth]{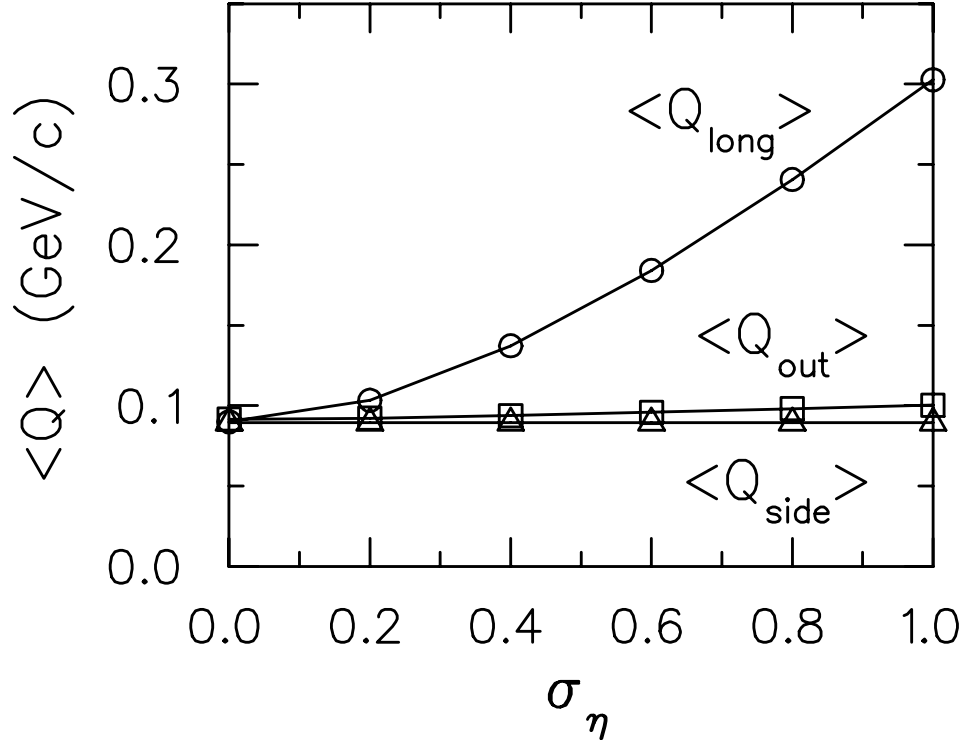}}
\caption{
\label{fig:sigmaeta}
The width of the balance function is shown for the three momentum
components. The calculations assumed a blast wave scenario with the collective
velocities of the source points for balancing pions being separated
longitudinally according to a Gaussian distribution of width $\sigma_\eta$. The
calculations assumed a breakup temperature of 120 MeV and a maximum transverse
collective velocity of 0.7$c$.
}
\end{figure}

The balance function could also be binned in any of the three projections,
$Q_{\rm long}$, $Q_{\rm out}$ and $Q_{\rm side}$, rather than in $Q_{\rm
inv}$. If the balancing pairs were to always originate from sources with the
same collective velocity, the balance function would be identical in all three
variables. However, if the balancing particles were to diffuse relative to one
another, the shape of the balance function might become decidedly
non-isotropic. For instance, if charge is created early in a RHIC collision,
the balancing charges might easily separate along the beam axis and ultimately
be emitted from regions with different rapidities. Figure \ref{fig:sigmaeta}
presents the widths of balance functions assuming that the balancing particles
independently dissipated and were each ultimately emitted with sources moving
with a spread of rapidities characterized by $\sigma_\eta$. A Gaussian form for
the diffusion was assumed for the distribution of source rapidities, $y_s$.
\begin{equation}
P(y_s)\sim \exp\left(\frac{y_s^2}{2\sigma_\eta^2}\right).
\end{equation}
This extends the distribution of $Q_{\rm long}$ while leaving the distribution
of $Q_{\rm side}$ unaffected and the distribution of $Q_{\rm out}$ only
slightly affected by boost effects. As can be seen in Fig. \ref{fig:sigmaeta},
the disparity in the three widths should be easily observed for this example
where the temperature was chosen to be 120 MeV and the radial collective
velocities were between zero and $v_{\rm max}=0.7c$.

It should be difficult to discern the difference between thermal broadening and
dissipation of balancing charges into regions with different collective
flow. However, other observables provide insight into breakup temperature,
mainly the comparison of proton and pion spectra
\cite{rhic_protonpionspectra}. Once one knows the breakup temperature, it is
possible to fit parameters that describe the diffusive spread,
e.g. $\sigma_\eta$. Furthermore, a thermal fit to data where the diffusive
terms are set to zero provides an upper bound for the breakup temperature.

For the reasons above, much of the analysis of the following sections will be
given in $Q_{\rm inv}$. An additional advantage of using $Q_{\rm inv}$ is that
it allows one to identify the contributions from specific resonances which
contribute peaks to the balance function when plotted in $Q_{\rm inv}$. It is
our hope that experimental analyses will also switch to these variables.

\section{The effects of surplus positive charge}
\label{sec:surplus}

Not all charges have balancing partners. In a Au+Au collision at RHIC, the two
gold nuclei provide 158 unbalanced protons and 236 unbalanced neutrons. These
pollute the balance function by providing unbalanced electric charge, baryon
number and isospin. For detectors like STAR, these effects are lessened by the
fact that most of the surplus charge is at high rapidity and outside the
experimental acceptance. However, the effect should become more significant if
the balance function is constructed for a set of charges, e.g. $p\bar{p}$, for
which there is a significant imbalance of one charge vs. the opposite
charge. Our goal in this section is to offer a revised procedure for producing
balance functions from data that would subtract the pollution due to the
surplus charge. More precisely, we wish to define a balance function that would
ignore any additional unbalanced charges that are not correlated with one
another or with pair-wise created charges.

In order to demonstrate the effects of the polluting surplus charge, we
introduce a notation where distributions, $\bar{N}$, count charges which are
divided into three categories. The subscripts ``+'' and ``-'' will refer to
positive and negative charges which are created in tandem. The subscripts
``$\delta$'' will denote the surplus positive charge. The balance function will
be re-evaluated after inserting the replacements,
\begin{equation}
\begin{split}
N_-\rightarrow& \bar{N}_-,\\
N_+\rightarrow& \bar{N}_+ +\bar{N}_\delta,\\
N_{--}\rightarrow& \bar{N}_{--},\\
N_{++}\rightarrow& \bar{N}_{++}+\bar{N}_{\delta +}
+\bar{N}_{+\delta}+\bar{N}_{\delta\delta},\\
N_{+-}\rightarrow&\bar{N}_{+-} +\bar{N}_{\delta -},\\
N_{-+}\rightarrow&\bar{N}_{-+}+\bar{N}_{-\delta},
\end{split}
\end{equation}
into Eq. (\ref{eq:balancedef}).
\begin{equation}
\label{eq:balancedef_wdelta}
\begin{split}
B(P_2|P_1)=&\frac{(2\bar{N}_+(P_1)+\bar{N}_\delta(P_1))}
{2(\bar{N}_+(P_1)+\bar{N}_\delta(P_1))}
\cdot\frac{\bar{N}_{+-}(P_1,P_2)-\bar{N}_{++}(P_1,P_2)}{\bar{N}_+(P_1)}\\
&+\frac{\bar{N}_\delta(P_1)\bar{N}_{+\delta}(P_1,P_2)
-\bar{N}_+(P_1)\bar{N}_{\delta\delta}(P_1,P_2)}
{2(\bar{N}_+(P_1)+\bar{N}_\delta(P_1))\bar{N}_+(P_1)}.
\end{split}
\end{equation}
In deriving Eq. (\ref{eq:balancedef_wdelta}) an explicit symmetry between the
positive and negative charges has been assumed, i.e., $\bar{N}_+=\bar{N}_-$,
$\bar{N}_{++}=\bar{N}_{--}$, and $\bar{N}_{+-}=\bar{N}_{-+}$. The first term in
this expression is proportional to the unpolluted balance function,
$\bar{B}$. The second term can be simplified by assuming that the surplus
charges are uncorrelated with the other charges and that they are also
uncorrelated with themselves, aside from overall conservation of charge.
\begin{equation}
\bar{N}_+(P_1) \bar{N}_{\delta\delta}(P_1,P_2)
=\bar{N}_\delta(P_1) \bar{N}_{\delta +}(P_1,P_2)
\frac{Q-1}{Q},
\end{equation}
where $Q$ represents the maximum integrated surplus charge. The balance
function can then be expressed as:
\begin{equation}
\label{eq:balancepollution}
\begin{split}
B(P_2|P_1)=&\frac{(2N_+(P_1)+\bar{N}_\delta(P_1))}{2(\bar{N}_+(P_1)
+\bar{N}_\delta(P_1))}\bar{B}(P_2|P_1)+\frac{1}{Q-1}\cdot
\frac{\bar{N}_{\delta\delta}(P_1,P_2)}
{2(\bar{N}_+(P_1)+\bar{N}_\delta(P_1))},\\
\bar{B}(P_2|P_1)\equiv&\frac{\bar{N}_{+-}(P_1,P_2)
-\bar{N}_{++}(P_1,P_2)}{\bar{N}_+(P_1)}.
\end{split}
\end{equation}
If the charges used to construct the balance function obey strict charge
conservation, a perfect detector would satisfy the normalization conditions,
\begin{equation}
\label{eq:normconditions}
\begin{split}
\sum_{P_2} \bar{B}(P_2|P_1)=&1,\\
\sum_{P_2}\bar{N}_\delta(P_2)=&Q,\\
\sum_{P_2}\bar{N}_{\delta\delta}(P_1,P_2)=&N_\delta(P_1)\frac{Q-1}{Q}.
\end{split}
\end{equation}
After inserting Eq. (\ref{eq:normconditions}) into
Eq. (\ref{eq:balancepollution}), one can see that the normalization of the
balance function is unchanged by the surplus charge. However, the shape is
altered as the balance function is comprised of two components. The first term
in Eq. (\ref{eq:balancepollution}) describes the separation of balancing
charges, while the second term is governed by the separation of two random
balancing charges. The relative weights of the two terms is determined by the
fraction of the charge which owes itself to a surplus in the initial
state. Thus, the effect of the surplus charge is to dampen the contribution
from the balancing charges and to average in a second contribution.

For the STAR detector at RHIC, this second term is fairly small even for
protons as the number of surplus protons is less than 10 per unit rapidity in
central collisions \cite{brahms_jhlee_qm2002}. Given that charge conservation
constraints would suggest $Q=158$, the effect of the extra charge is to first
dampen the balance function by approximately 15\%, and secondly to add in a
second component whose width is characteristic of the acceptance, and whose
magnitude is only one or two percent of the contribution from balancing
charges.

In order to eliminate the contribution from surplus charge and determine
$\bar{B}$ from experiment, one can consider an object similar as to what is
used to create the balance function numerator using mixed events. This object
will be referred to as $M(P_1,P_2)$ and will be constructed from mixed events,
where $N^m_{a,b}(P_1,P_2)$ signifies that the charges $a$ and $b$ which satisfy
the momentum constraints $P_1$ and $P_2$ are chosen from separate events.
\begin{equation}
\begin{split}
\label{eq:mixeddef}
M(P_1,P_2)\equiv& N^m_{+-}(P_1,P_2)-N^{m}_{++}(P_1,P_2)+N^m_{-+}(P_1,P_2)
-N^m_{--}(P_1,P_2)\\
=&\bar{N}^m_{+-}(P_1,P_2)-\bar{N}^{m}_{++}(P_1,P_2)+\bar{N}^m_{-+}(P_1,P_2)
-\bar{N}^m_{--}(P_1,P_2)\\
&-\bar{N}^m_{+\delta}(P_1,P_2)-\bar{N}^m_{\delta +}(P_1,P_2)
-\bar{N}^m_{\delta\delta}(P_1,P_2)
+\bar{N}^m_{-\delta}(P_1,P_2)+\bar{N}^m_{\delta -}(P_1,P_2).
\end{split}
\end{equation}
Since the counts for different events are independent,
$N^m_{+-}=N^m_{++}=N^m_{-+}=N^m_{--}$, $N^m_{\delta +}=N^m_{\delta -}$ and
$N^m_{+\delta}=N^m_{-\delta}$. Thus, $M$ becomes
\begin{equation}
M(P_1,P_2)=-\bar{N}^m_{\delta\delta}(P_1,P_2).
\end{equation}
One could define a similar object using pairs from the same event,
\begin{equation}
\begin{split}
N(P_1,P_2)\equiv&N_{+-}(P_1,P_2)-N_{++}(P_1,P_2)+N_{-+}(P_1,P_2)
-N_{--}(P_1,P_2)\\
=&\bar{N}_{+-}(P_1,P_2)-\bar{N}_{++}(P_1,P_2)+\bar{N}_{-+}(P_1,P_2)
-\bar{N}_{--}(P_1,P_2)-\bar{N}_{\delta\delta}(P_1,P_2).
\end{split}
\end{equation}
Again if one assumes that the only correlation between surplus charges involves
a constraint on the overall number of surplus charges,
\begin{equation}
N(P_1,P_2)=
\bar{N}_{+-}(P_1,P_2)-\bar{N}_{++}(P_1,P_2)+\bar{N}_{-+}(P_1,P_2)
-\bar{N}_{--}(P_1,P_2)-\frac{Q-1}{Q}\bar{N}_{\delta\delta}^m(P_1,P_2).
\end{equation}
One can then generate the desired balance function $\bar{B}(P_2|P_1)$ by
combining $N$ and $M$,
\begin{equation}
\bar{B}(P_2|P_1)=\frac{N(P_1,P_2)-M(P_1,P_2)(Q-1)/Q}{N_<(P_1)},
\end{equation}
where $N_<$ is $N_-$ if the surplus charges are positive and is $N_+$ if the
surplus charges are negative.

It should be emphasized that this derivation assumed that the detector has
equal acceptance for positive and negative charges. The ability of $\bar{B}$
to ignore the polluting surplus charge is based on the assumption that the
surplus charges are uncorrelated with one another and are equally
correlated with the ``+'' and ``-'' charges. As discussed in the next section
and in \cite{jeonpratt}, such correlations can be important, especially at
small relative momentum.

\begin{figure}
\centerline{\includegraphics[width=0.7\textwidth]{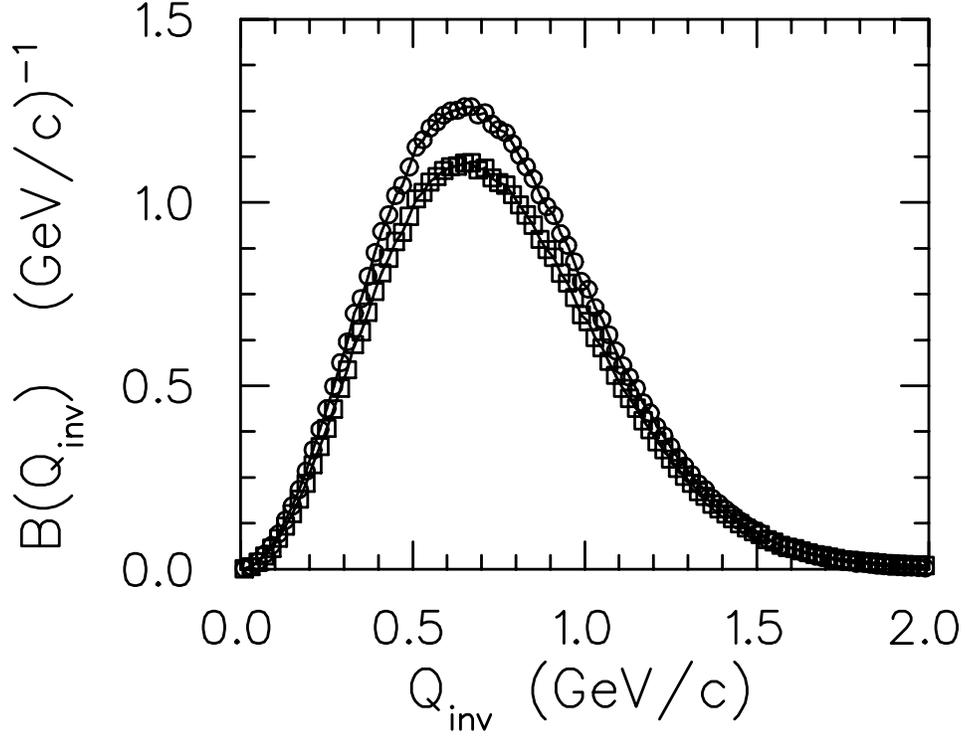}}
\caption{
\label{fig:surplus}
Proton-antiproton balance functions are shown for a blast wave model with and
without corrections for the surplus unbalanced protons. The corrected balance
function (circles) is constructed assuming a breakup temperature of 120 MeV, a
maximum transverse velocity of 0.7$c$, and $\sigma_\eta=0$. The calculation was
scaled down by 40\% to account for balancing of the proton's charges by other
species. The distorted balance function (squares) is based on the proton excess
as measured by the BRAHMS collaboration. Both balance functions were filtered
through the STAR acceptance.  }
\end{figure}

To illustrate the importance of these corrections, we generate a $p\bar{p}$
balance function from a simple model of a boost-invariant emission of particles
governed by a temperature of 120 MeV and a maximum transverse velocity of
0.7$c$. It is assumed that the number of protons per unit rapidity is 28 and
that the number of antiprotons is 21 to be consistent with measurements from
RHIC \cite{brahms_jhlee_qm2002}. The polluted balance function as described in
Eq. (\ref{eq:balancepollution}) is displayed in Fig. \ref{fig:surplus} along
with the corrected balance function $\bar{B}$. This calculation is generated by
assuming that particles were emitted from sources with random rapidities, but
that two balancing particles are emitted from sources with the same
velocity. The parameter $Q$ used in Eq. (\ref{eq:balancepollution}) is assumed
to be 158.

Two additional modifications have been added to Eq. (\ref{eq:balancepollution})
in order to more fairly illustrate the magnitude of the effect of the surplus
charge. First, the function $\bar{B}$ was scaled down by 40\% to account for
the fact that the charge of an antiproton is often balanced by a neutron or by
a $\Lambda$. Secondly, the simulated momenta were put through an acceptance
filter which crudely mocks the acceptance of the STAR detector at
RHIC. Particles were required to have a $p_t$ greater than 100 MeV/c and a
momentum of magnitude less than 700 MeV/c. The pseudo-rapidities were confined
to a region of midrapidity, $-1.1<\eta<1.1$.

As illustrated in Fig. \ref{fig:surplus}, the effects of the extra charge are
mainly to dampen the balance function. The importance of correcting for the
surplus charge would certainly be magnified if one were to analyze balance
functions from SPS or AGS collisions where the fraction of extra protons is
much higher. These corrections are not-model dependent, and the corrected
balance functions exactly reproduce $\bar{B}$. However, it should be emphasized
that this statements relies on the assumption that the surplus charge is
uncorrelated with other surplus charges, and with the pair-wise created
charges.

\section{Final-State Interaction Distortion to the Balance Function}
\label{sec:hbt}
The balance function is implicitly predicated on the assumption that there are
no residual correlations between a given charge and all other charges besides
its balancing partner, i.e., all other charges are statistically eliminated
from the distribution by the like-sign subtraction. Not all correlations cause
problems. For instance, flow correlations tend to be identical between
particles of the opposite charge or the same charge and thus fall out of the
balance function. On the other hand final-state interactions involve all the
other charges and depend sensitively on the relative signs of the charges. This
distortion can rise linearly with the multiplicity since the number of charges
with which a given charge can correlate rises linearly with the
multiplicity. However, correlation functions tend to approach unity at higher
multiplicity in accordance with expectations for increasing source size.  This
makes the resulting multiplicity dependence of the distortion non-trivial.

A method for estimating the distortion to the balance function from residual
interactions was provided in \cite{jeonpratt}. The same method is applied
here. For every balancing pair, $p_a$ and $p_b$, one must consider the
correlation weight with other pairs whose momenta are $p_c$ and $p_d$. The
weight, $w(p_a,p_b;p_c,p_d)$ can be estimated:
\begin{equation}
\label{eq:cweight}
w(p_a,p_b;p_c,p_d)\approx C_{++}(p_a,p_c)C_{--}(p_b,p_d)
C_{+-}(p_a,p_d)C_{-+}(p_b,p_c).
\end{equation}
Ideally, the balance function would isolate the $ab$ pair and the interaction
with the $cd$ pair would cancel from the subtraction, $N_{+-}-N_{++}$. The
correlations will lead to distortions if
\begin{equation}
w(p_a,p_b;p_c,p_d)\ne w(p_a,p_b;p_d,p_c).
\end{equation}
I.e., distortions are caused only by those interactions which differ between
same-sign and opposite-sign particles. For instance, an isoscalar exchange of
between pions would not bring on a distortion, but a Coulomb interaction or
identical-particle interference would provide a source for distortion.

We simulate these effects for $\pi^+\pi^-$ balance functions with the same
blast wave model described in Section \ref{sec:qinv}. The source of the $ab$
pairs was chosen to move with one randomly chosen velocity, while the $cd$
source was chosen to move with a different velocity. The distributions were
calculated for the balance function numerators using the four particles, but
rather than incrementing he distributions by unity, the distributions were
incremented by the weight described in Eq. (\ref{eq:cweight}). By considering
the contributions due to the extra weight separately from the usual
contributions arising between the balancing particles, and weighting them
appropriately for the given value of $dn/dy$, balance functions were
calculated with and without the distortions.

In addition to the blast-wave parameters, the residual-interaction distortion
is sensitive to the form for the two particle correlations and to the pion
multiplicity. The multiplicity of charged pions was chosen to be 300 for both
positive and negative pions, roughly consistent with measurements at RHIC
\cite{rhicpionmultref}. The correlation functions were chosen to correspond to
Gaussian sources of radius, $R_{\rm inv}=6.0$ fm, again consistent with
measurements at RHIC \cite{starhbt}. A more sophisticated treatment would
account for the $p_t$ dependence and the directional dependence of the
radius. The correlation calculation presented here accounted for both
identical-particle interference and for the mutual Coulomb interaction.

\begin{figure}
\centerline{\includegraphics[width=0.7\textwidth]{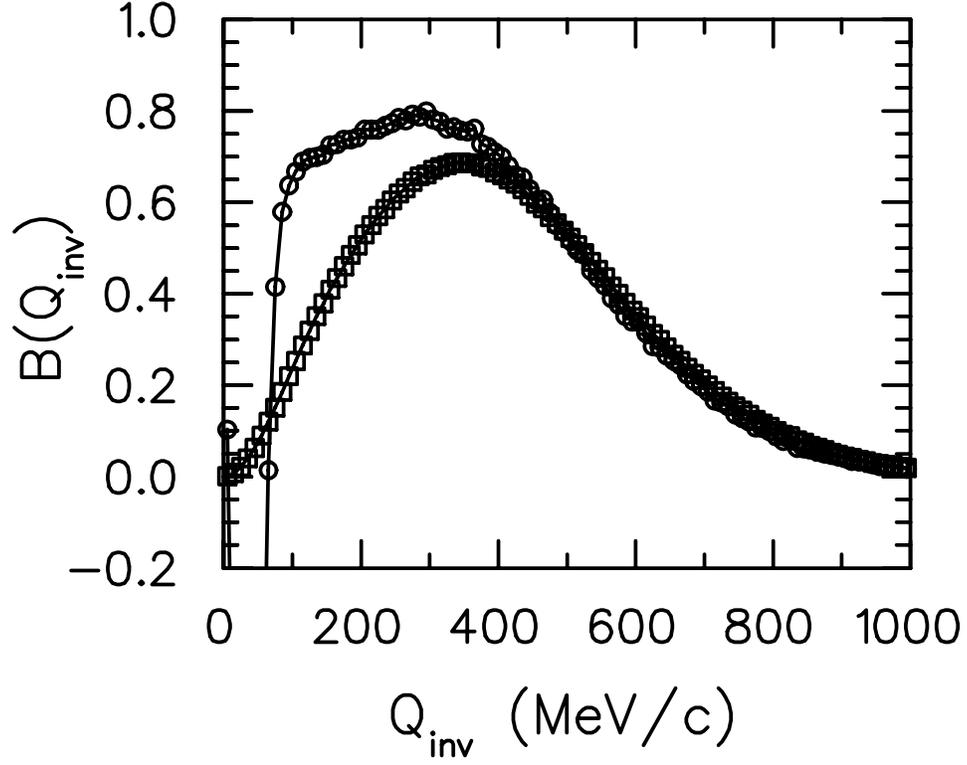}}
\caption{
\label{fig:balancehbt_qinv}
As a function of $Q_{\rm inv}$, $\pi^+\pi^-$ balance functions from a
blast-wave model are shown with (circles) and without (squares) the distorting
effects of inter-pair interactions. The model assumed a breakup temperature of
120 MeV, a maximum transverse velocity of 0.7$c$, and $\sigma_\eta=0$. The
undistorted balance function was scaled by 70\% to account for balancing by
other species, and both balance functions were filtered by the STAR
acceptance. The significant enhancement for momenta between 60 MeV/c and 400
MeV/c owes itself to the Coulomb interaction between pions. }
\end{figure}

\begin{figure}
\centerline{\includegraphics[width=0.7\textwidth]{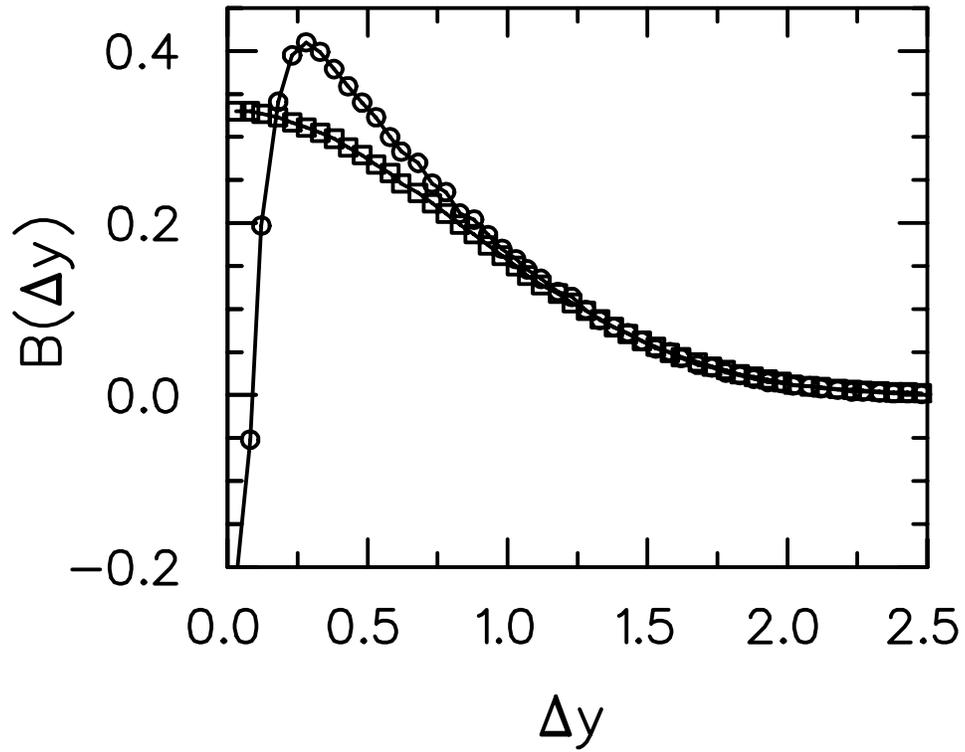}}
\caption{
\label{fig:balancehbt_dely}
The same as Fig. \ref{fig:balancehbt_qinv}, only with the balance function
being plotted as a function of relative rapidity. The distorting effects are
less noticeable in $\Delta y$.
}
\end{figure}

The results of the calculation are shown in Fig.s \ref{fig:balancehbt_qinv} and
\ref{fig:balancehbt_dely}. A crude filter for the STAR detector at RHIC was
applied and it was assumed that 70\% of the pions had their charge balanced by
other pions, which affects the normalization of the balance function.

Residual interactions can either strengthen or diminish the balance function
depending on the relative momentum. At very small relative momentum, the
balance function at small relative momentum rises due to the Coulomb
enhancement of the $\pi^+\pi^-$ correlation function and the Coulomb repulsion
of the $\pi^+\pi+$ and $\pi^-\pi^-$ correlation functions. For values of
$Q_{\rm inv}$ larger than a few MeV/c but less than $\sim 25$ MeV/c, the
identical-particle interference which enhances the $\pi^+\pi^+$ and
$\pi^-\pi^-$ correlation functions, diminishes the balance function since
same-sign pairs contribute negatively to the balance function. At larger
relative momenta, Coulomb effects again dominate. The effects are less dramatic
when the balance function is viewed as a function of relative rapidity.

\begin{figure}
\centerline{\includegraphics[width=0.7\textwidth]{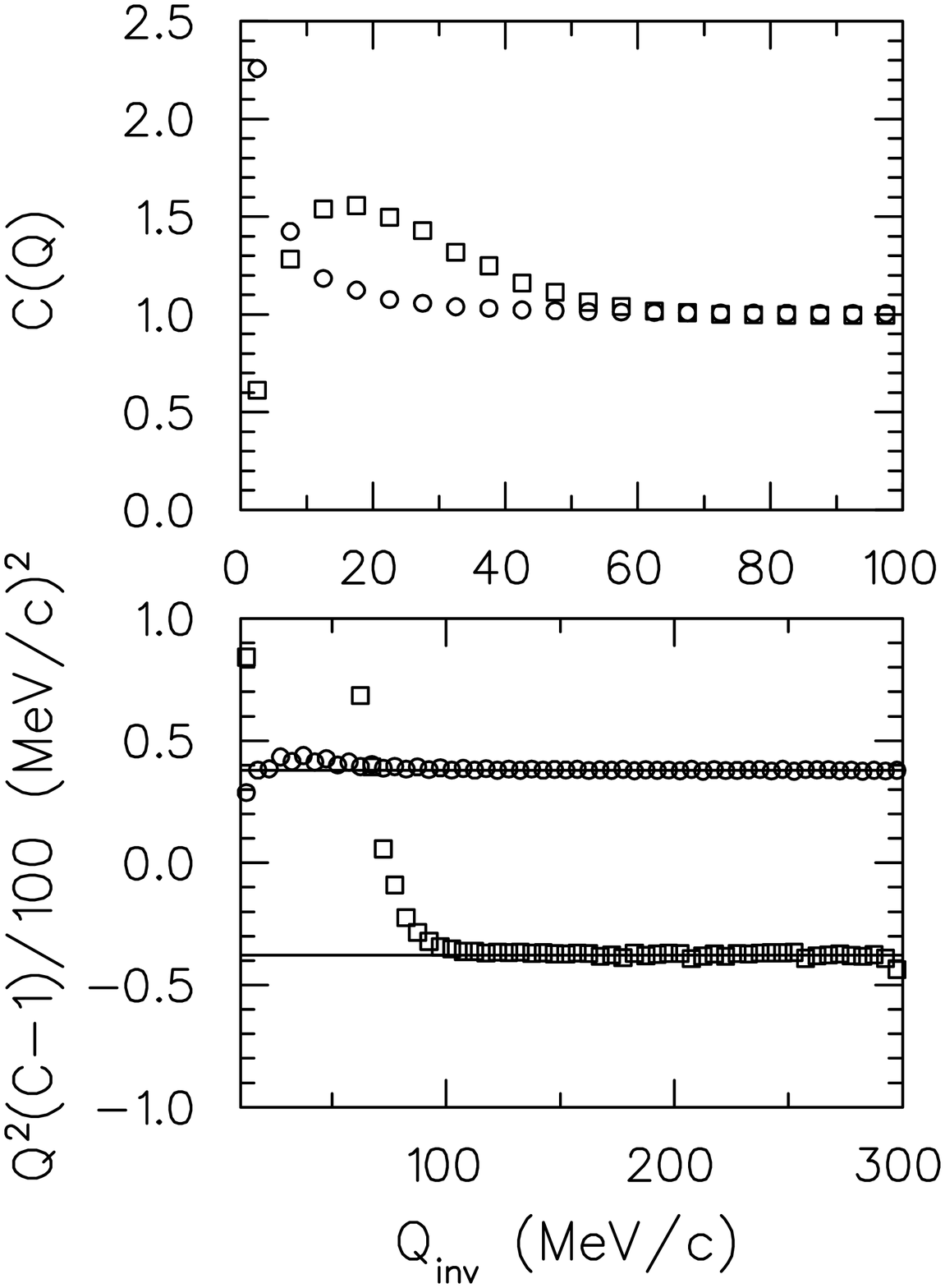}}
\caption{
\label{fig:corrtail}
Correlations for same-sign (squares) and opposite-sign (circles) pions are
shown for a Gaussian source of size $R=6$ fm in the upper panel. To illustrate
the Coulomb effects that
force $C(Q)$ to approach unity as $1/Q^2$, $C(Q)-1$ is multiplied by $Q^2$ and
displayed in the lower panel. The lines represent the constant expected for
Coulomb interactions described in Eq. (\ref{eq:corrtailgauss}). Since phase
space increases as $Q^2$, inter-pair correlations distort the balance functions
for relative momenta of several hundred MeV/c.
}
\end{figure}

The distortion of the balance function in Fig. \ref{fig:balancehbt_qinv} is
dominated by Coulomb effects at large momentum. The correlation weights are
driven by the squared quantum wave function. However, the correlation for
large values of $qR$ can be understood by considering the classical analog to
the wave function. As shown in Ref. \cite{kimpratt}, the classical analog is:
\begin{equation}
\label{eq:corr_class_approx}
\begin{split}
\phi^2(q,r)\rightarrow& \frac{d^3q_i}{d^3q_f}=\frac{q_i}{q_f}\\
=&\sqrt{1-8Z_qZ_b\alpha\mu/rq_f^2}\\
\approx&1-4Z_aZ_b\alpha\mu/rq_f^2,
\end{split}
\end{equation}
where $\alpha$ is the fine structure constant, the product of the charges of
the two species is $Z_aZ_b$, $\mu$ is the reduced mass and $q_i$ and
$q_f=Q_{\rm inv}$ are the initial and final relative momenta. Thus, the effects
of Coulomb interactions, in the classical limit, only diminish as a function of
$1/Q_{\rm inv}^2$. By averaging $1/r$ over a Gaussian source characterized by
the Gaussian source size $R$, one can find the asymptotic form for the
classical correlation function.
\begin{equation}
\label{eq:corrtailgauss}
C_{\rm class}(Q_{\rm inv})\approx 1-\frac{4\mu\alpha Z_aZ_b}
{Q_{\rm inv}^2R\sqrt{\pi}}.
\end{equation}
The classical result for the $\pi^+\pi^-$ correlation function is compared to
the quantum result for a six-fm source in Fig. \ref{fig:corrtail}. The
agreement is remarkable for $Q_{\rm inv}>25$ MeV/c when $qR>1$, especially for
the opposite-sign case where there is no identical-particle interference.

Even though the correlation function goes to zero proportional to $1/Q_{\rm
inv}^2$, the phase space is increasing as $Q_{\rm inv}^2$. Thus, the Coulomb
interaction remains important to remarkably large momenta. For more central
collisions, the value of $R$ rises, but the number of particles with which a
given particle is correlated also rises. If the multiplicity scales as $R^3$,
it is clear that the Coulomb distortion will become acute for central
collisions.

Strong-interaction distortions have not been considered in these
calculations. In terms of the $\pi\pi$ phase shifts, $\delta_\ell$, the
contribution to the correlation function from strong interactions can be
approximated by the relation \cite{jenningsboalshillcock},
\begin{equation}
C(q)\sim \frac{1}{4q^2R^3\sqrt{\pi}}(2\ell+1)\frac{d\delta_\ell}{dq}.
\end{equation}
The strength of the strong-interaction correlation falls much more quickly with
$R$ than does Coulomb-induced correlation. If $R^3$ were to scale linearly with
multiplicity, the effect of the strong interaction on the balance function
would be roughly independent of multiplicity or centrality. This difference in
the behavior derives from the fact that a given pion interacts with only its
neighbors through the strong interaction while it may interact with nearly all
particles through the Coulomb interaction. If the breakup density is
independent of centrality, the number of neighbors stays constant and the
distortion to the balance function from strong interactions should not be
strongly centrality dependent. On the other hand, the Coulomb distortion
interaction should be much stronger for central collisions than for peripheral
collisions.

The ingredients for calculating the distortion were the correlation weights and
spectra along with the procedure for generating the pairs, $(p_a,p_b)$ and
$(p_c,p_d)$. In principal, the spectra and correlation weights can be taken or
inferred from data without introducing a theoretical model. However, the
generation of the pairs can not be extracted directly from data due to the
correlations between $p_a$ and $p_b$ and those between $p_c$ and $p_d$. Since
particles are produced pairwise, it is necessary to include these correlations
because the inter-pair interaction must attract pairs rather than single
particles. I.e., the net charge in the medium can not change, but it can be
polarized. At face value, this is an explicit model dependence. However, the
parameters that govern the correlation between $a$ and $b$ and between $c$ and
$d$ are precisely those parameters used to model the undistorted balance
function. Thus, no additional model parameters would be introduced to calculate
the distortion. Thus, the distortion from residual interactions can not be
subtracted from experimental results in a model-independent fashion, but it can
be modeled theoretically without additional parameters.

Given the significant effects from inter-pair correlations, it is imperative
that the balance function analyses correct for these distortions.  Fortunately,
the corrections can be confidently modeled, and the robustness of the balance
function is not compromised. However, this conclusion is predicated on an
understanding of the two-particle correlations. Since a correlation of a
fraction of a percent can significantly alter the balance function, the issue
of strong-interaction corrections to the balance functions should be
revisited. 

Strong-interaction effects can be divided into two categories. The first
category would be $s$-channel interaction which have particle-antiparticle
channels, e.g. $\rho^0\rightarrow \pi^+\pi^-$. But, this source should not be
considered as a distortion since the two pions are indeed a balancing pair. For
instance, if all pions resulted from $\rho^0$ decays, the balance function
would peak at the invariant mass of the $\rho$, and provide an important clue
as to the creation mechanism for pions. Such resonant contributions can be
calculated in a microscopic model or in a thermal calculation based on the
canonical ensemble. A second source of strong-interaction effects is the
interaction with other bodies through non-resonant interactions. Since the
strong interaction is short range, this interaction should involve only a few
neighbors. For large sources, the Coulomb interaction provides a larger effect
on two-particle correlation function than does the strong
interaction. Nonetheless, it would be worthwhile to better quantify the
significance or insignificance of the strong interaction.

\section{Summary and Discussion}
\label{sec:summary}
Charge balance functions were developed with the hope of identifying balancing
charges on a statistical basis. Two effects prevent the like-sign subtraction
from accomplishing this goal to high precision. As shown in
Sec. \ref{sec:surplus}, the excess nucleons coming from the colliding nuclei
provide a modest pollution to the balance function in measurements at
mid-rapidity at RHIC. Fortunately, these effects can be easily subtracted. The
second source of distortion derives from the inter-pair interaction of
balancing charges with other particles in the medium. These distortions become
more important in high multiplicity events. As shown in Sec. \ref{sec:hbt}, for
high multiplicity events these distortions are most strongly affected by the
Coulomb interaction. These effects are also more noticeable for balance
functions calculated in $Q_{\rm inv}$ than they are for balance functions
calculated in relative rapidity. Although it is difficult to subtract these
distortions in a model-independent fashion, it is straight-forward to include
these effects in a theoretical treatment. In addition to the typical parameters
one would use to model balance functions, modeling the distortion requires only
an additional understanding of two-particle correlations. As these correlations
can be extracted from measurement, the distortion from inter-pair interactions
can be modeled quite confidently. In central collisions, this distortion can be
a 20\% effect, and if the correlation functions are understood to the 90\%
level, the residual systematic uncertainty is probably of the order of one or
two percent.

\begin{acknowledgments}
This work was supported by the National Science Foundation, Grant No.
PHY-00-70818.
\end{acknowledgments}

\begin{thebibliography}{99}
\bibitem{bassdanpratt} S.A. Bass, P. Danielewicz and S. Pratt, Phys. Rev.
Lett. {\bf 85}, 2689 (2000).
\bibitem{jeonpratt} S. Jeon and S. Pratt, Phys. Rev. C{65}, 044902 (2002).
\bibitem{asakawaprl} M. Asakawa, U. Heinz and B. Muller,
  Phys. Rev. Lett. {\bf 85}, 2072 (2000).
\bibitem{asakawa} M. Asakawa, U.W. Heinz and B. Muller, Nucl. Phys. A{\bf 698},
519 (2002).
\bibitem{kochpolonica} V. Koch, Acta Phys. Polon. B{\bf 33}, 4219 (2002).
\bibitem{kochqm2001} V. Koch, M. Bleicher and S. Jeon, Nucl. Phys. A{\bf 698},
  261 (2002).
\bibitem{bleicherjeonkoch} M. Bleicher, S. Jeon and V. Koch, Phys. Rev. C{\bf
  62}, 061902 (2000).
\bibitem{jeonkoch} S. Jeon and V. Koch, Phys. Rev. Lett. {\bf 85}, 2076 (2000).
\bibitem{gavinqm2002} M. Abdel-Aziz and S. Gavin, Proceedings of Quark Matter
  2002, to appear in Nucl. Phys. A.
\bibitem{pruneaugavinvoloshin} C. Pruneau, S. Gavin and S. Voloshin,
  Phys. Rev. C{\bf 66}, 044904 (2002).
\bibitem{gavinkapusta} S. Gavin and J.I. Kapusta, Phys. Rev. C{\bf 65}, 054910
  (2002). 
\bibitem{ray_qm2002star} L. Ray (STAR collaboration), Proceedings of
  Quark Matter 2002, Nantes, France, to appear in Nucl. Phys. A.
\bibitem{prattprd86} S. Pratt, Phys. Rev. D{\bf 33}, 1314 (1986).
\bibitem{bertschoutlongside} G.F. Bertsch, Nucl. Phys. A{\bf 498}, 173c
(1989).
\bibitem{csorgopratt} T. Cs\"{o}rg\H{o}, J. Zim\'{a}nyi, J. Bondorf, 
H. Heiselberg, and S. Pratt, Phys. Lett. {\bf B241}, 301 (1990).
\bibitem{rhic_protonpionspectra} K. Adcox, et al., Phys. Rev. Lett. {\bf 88},
  242301 (2002).
\bibitem{brahms_jhlee_qm2002} J.H. Lee (BRAHMS collaboration), Proceedings of
  Quark Matter 2002, Nantes, France, to appear in Nucl. Phys. A.
\bibitem{rhicpionmultref} T. Chujo (PHENIX collaboration), Proceedings of Quark
  Matter 2002, Nantes, France, to appear in Nucl. Phys. A. 
\bibitem{starhbt} C. Adler, et al., Phys. Rev. Lett. {\bf 87}, 082301 (2001).
\bibitem{kimpratt} Y.D. Kim, R.T. de Souza, C.K. Gelbke, W.C. Gong and
  S. Pratt, Phys. Rev. C{\bf 45}, 387 (1992).
\bibitem{jenningsboalshillcock} B.K. Jennings, D.H. Boal and J.C. Shillcock,
Phys. Rev. C{\bf 33}, 1303 (1986).
\end{thebibliography}
\end{document}